\documentclass[%
 reprint,
showpacs,preprintnumbers,
 amsmath,amssymb,
 aps,
]{revtex4-1}

\usepackage{graphicx}
\usepackage{dcolumn}
\usepackage{bm}

\begin{document}

\title{Quantum-information Division and an Optimal Uncorrelated Channel}

\author{Yuji Sekino}
 
\author{Satoshi Ishizaka}%

\affiliation{Graduate School of Integrated Arts and Sciences, Hiroshima University, 
 Higashi-Hiroshima 739-8521, Japan}%


\date{\today}

\begin{abstract}
We consider {\it quantum-information division}, which is characterized by a channel 
whose outputs have no correlation and are not completely randomized.
We show that the quantum-information division is possible in a probabilistic manner
by optimizing the average fidelity in the channel with $M$ outputs
in both deterministic and probabilistic
cases. Moreover, we show that the optimal fidelities drastically change depending on the condition imposed on the outputs (symmetric and asymmetric), which is quite in contrast to the case of imperfect cloning. 
\end{abstract}
\maketitle


\section{Introduction}
The no-cloning theorem \cite{bookb, bookc}, which states that an unknown quantum state 
cannot be perfectly copied, is a cornerstone of quantum physics.
In spite of the no-go theorem, however, it was shown by Bu$\check{z}$ek and Hillery \cite{buhi}
that imperfect cloning, where average fidelity between 
the original unknown state (input) and the copied state (output) does not reach $1$, is possible.
After their insight, the 
Bu$\check{z}$ek-Hillery imperfect quantum cloning machine was proved to be optimal
in the sense of average fidelity \cite{gima,brdi,gi}. 
For an intensive review of quantum cloning, see Ref. \cite{scib}. 

At first glance, imperfect cloning seems to divide unknown quantum information of the input into
the outputs.
As we can see in the Bu$\check{z}$ek-Hillery imperfect quantum cloning machine for instance, however,
there remains correlation among the output states in general.
Namely, the output states are no longer independent of each other, 
and thus quantum information is regarded to be 
distributed rather than divided among the outputs in imperfect cloning. In order to say 
that "quantum information is divided," it would be at least necessary that there exists no correlation including not only quantum correlation such as entanglement but also classical one
among the outputs. Namely, the output state  of cloning with $M$ outputs for each input state $|\psi\rangle$ should have the form of $\rho_1 (\psi)\otimes  \rho_2 (\psi)\otimes \cdots \otimes \rho_M (\psi)$, where $\rho_i (\psi)$ is the $i$-th output as a function of $|\psi\rangle$.
Now we have 
a question: Is imperfect cloning without correlation among outputs possible, or is {\it uncorrelated cloning} possible?
This is not a trivial issue.
For instance, let us consider the cloning strategy called
{\it the measurement-based procedure} \cite{scib}, where an input state is measured in the $\{\vert 0\rangle, \vert 1\rangle \}$ basis, and either $\vert 0\rangle\vert 0\rangle$
or $\vert 1\rangle\vert 1\rangle$ is prepared depending on the measurement outcomes. This seems to achieve the desired uncorrelated cloning, but for the input state of 
$\vert \psi\rangle =\sqrt{p}\vert 0\rangle +\sqrt{1-p}\vert 1\rangle$ ($0\leq p\leq 1 $ ), the procedure results in the output state of $p\vert 00\rangle\langle 00\vert +(1-p)\vert 11\rangle\langle 11\vert$. Clearly, the two outputs are classically correlated unless $p=0,1$, and hence this procedure does not achieve the uncorrelated cloning. 

Recently
D'Ariano {\it et al}. studied this issue from the viewpoint of {\it quantum-state decorrelation} \cite{dade}.
To be surprised the answer is negative, which means that no matter how small,
multiple outputs cannot depend on the  
input state $\vert \psi\rangle$ simultaneously
if no correlation among outputs is allowed;
only one output can depend on $\vert \psi\rangle$.
However, the quantitative evaluation by means of average fidelity is
crucial to see whether unknown quantum information can be divided or not.  
In this paper we study this issue by 
optimizing average fidelity in the information division into $M$ outputs in $d$-dimensional systems
in both deterministic and probabilistic cases. In each case,
the optimization is performed by imposing symmetric or asymmetric condition
on the output states. 
From the derived optimal values for each output, we conclude that quantum
information can be divided into multiple outputs probabilistically,
i.e., probabilistic {\it quantum-information division} is possible,
contrary to the previous suggestion \cite{dade}.    

This paper is organized as follows. 
In Sec. \ref{secqi}, we first rigorously define {\it quantum-information division}
by introducing an uncorrelated channel,
which can be expressed 
by means of average fidelity in the channel.   
In Sec. \ref{secou}, we optimize uncorrelated channels by means of entanglement fidelity,
and the resultant optimal average fidelities are shown. Finally, a summary and conclusion are given in 
Sec. \ref{con}.
In the Appendix,
we derive the state isomorphic to uncorrelated channel from the perspective of 
the Choi-Jamio\l kowski isomorphism, which is used for the optimization in Sec. \ref{secqi}.

\section{Quantum-information division}\label{secqi}
\newcommand{\1}{\mbox{1}\hspace{-0.25em}\mbox{l}}
Throughout this paper, we deal with the issue of dividing an unknown pure state in general.
So we consider a non-trace-preserving (trace-decreasing) channel,
which takes any $d$-dimensional pure state $\vert\psi\rangle$ as an input, and outputs  
$M$ states each with the same dimension $d$.  
Moreover, when the output states always do not have any correlation, we call it an uncorrelated channel. (Note that an uncorrelated channel excludes even 
classical correlation among outputs.)
Thus a map of the uncorrelated channel from an input to the $i$-th output can be written as
\begin{eqnarray}
	\vert \psi\rangle  \stackrel{\Lambda_{s_{i}}}{\longrightarrow}
    p_{_{\psi}}\rho _{s_{i}}(\psi ), \label{qia}
\end{eqnarray} 
where $p_{_{\psi}}$ is probability to output a state $\rho (\psi )$ and the suffix $S_{i}$ stands for
the $i$th-output system. 

In a trace-decreasing channel the output is realized with 
the probability $p_{_{\psi}}$ for input state $\vert\psi\rangle$.
So the average fidelity of a trace-decreasing channel $\Lambda_{s_{i}}$ is defined with the weight of 
probability $p_{_{\psi}}$ as
\begin{eqnarray}
	F_{s_{i}}
     =\frac{\displaystyle\int p_{_{\psi}}\langle\psi\vert\rho _{s_{i}}(\psi )\vert\psi\rangle d\psi}
	           {\displaystyle\int p_{_{\psi}}d\psi},	\label{pavfi}          
\end{eqnarray}	 
where the integral is over the uniform measure $d\psi$ on pure input states.
This definition is the natural generalization of the average fidelity of a 
trace-preserving channel \cite{ho}. 
When the average fidelity is $1/d$, the input and output are independent of each other.
Conversely, if average fidelity is not $1/d$, the output has some sort of information
of the input (i.e., not randomized) . 
Therefore, it can be defined such that {\it quantum-information division} is possible
if and only if $F_{s_{i}}\ne 1/d$ for all outputs $S_{i}$ in an uncorrelated channel.  

By the way, it is known as the Choi-Jamio\l kowski isomorphism
that a channel $\Lambda_{s_{i}}$ is completely characterized by a state  
$A_{_{RS_{i}}}=\lbrack I\otimes\Lambda_{s_{i}}\rbrack P^{+}_{_{RS_{i}}}$,
where $I$ is an identity, $P^{+}_{_{RS_{i}}}=\vert \Phi^{+}_{_{RS_{i}}}\rangle \langle\Phi^{+}_{_{RS_{i}}}\vert$
(\ $\vert \Phi^{+}_{_{RS_{i}}}\rangle =1/\sqrt{d}
\sum_{i=0}^{d-1}\vert i_{_{R}}\rangle\vert i_{_{S_{i}}}\rangle$)
and the suffix $R$ indicates a reference system used for
inputting half of the maximally entangled state into the system $S_{i}$.
The entanglement fidelity 
corresponding to a trace-decreasing channel $\Lambda_{s_{i}}$ is defined as
\begin{eqnarray}
	F^{e}_{s_{i}}=\hbox{Tr}\biggl[ P^{+}_{_{RS_{i}}}\ \displaystyle\frac{A_{_{RS_{i}}}}{\hbox{Tr}\lbrack A_{_{RS_{i}}}\rbrack}\biggr ].\label{penfi}          
\end{eqnarray}	 
Note that $\hbox{Tr}\lbrack A_{_{RS_{i}}}\rbrack$ is equal to the average probability
$\bar{p}=\int p_{_{\psi}}d\psi$ because, from the definition of $A_{_{RS_{i}}}$,
$\hbox{Tr}\lbrack A_{_{RS_{i}}}\rbrack=\hbox{Tr}\lbrack\Lambda_{s_{i}}(\1/d)\rbrack$  
holds and the integral is over the uniform measure $d\psi$ on input states. 

The channel isomorphic to the state $A_{_{RS_{i}}}/\hbox{Tr}\lbrack A_{_{RS_{i}}}\rbrack$
is obviously given by 
$\Lambda'_{s_{i}}=\Lambda_{s_{i}}/\hbox{Tr}\lbrack A_{_{RS_{i}}}\rbrack$. Then
the integral $\int \langle\psi\vert\Lambda'_{s_{i}}(\psi)\vert\psi\rangle d\psi$ coincides with the 
definition of average fidelity in a trace-decreasing channel (\ref{pavfi}). 
On the other hand, since $A_{_{RS_{i}}}/\hbox{Tr}\lbrack A_{_{RS_{i}}}\rbrack$ is a positive operator with unit trace, following the argument
by Horodecki {\it et al}. \cite{ho} (the channel with unit-trace isomorphic state becomes 
trace-preserving by twirling), the simple formula
\begin{eqnarray}
	F=\frac{d F^{e} +1}{d+1}	\label{paven}          
\end{eqnarray}	 
holds even if $F$ and $F^{e}$ are defined as Eq. (\ref{pavfi}) and (\ref{penfi}), respectively.

\section{Optimizing uncorrelated channel}\label{secou}
To evaluate channels how much input information can be divided into outputs, 
we consider the average of $F_{s_{i}}$ for all outputs $S_{i}$ ($i=1, 2,\cdots , M$)
\begin{eqnarray}
	\bar{F}=\displaystyle\frac{1}{M}\sum_{i=1}^{M}F_{s_{i}},\label{oaf}
\end{eqnarray}
and derive the optimal value of $\bar{F}$ under the uncorrelated condition.
Owing to the relation (\ref{paven}), we may optimize the average entanglement fidelity
$\bar{F}^{e}=\frac{1}{M}\sum_{i=1}^{M}F^{e}_{s_{i}}$
instead of $\bar{F}$. 
If we impose the uncorrelated condition to $A _{_{RS_{1}S_{2}\cdots S_{M}}}$, 
it must have the form of
\begin{eqnarray}
 A _{_{RS_{1}S_{2}\cdots S_{M}}}=X _{_{RS_{k}}} \otimes\bigotimes _{l\ne k}^{M} Y _{_{S_{l}}}^{l} \label{okk}
\end{eqnarray} 
for a certain $k$. This form is well expected from the result of \cite{dade}, but the rigorous proof is
given in the Appendix.
Therefore, the average entanglement fidelity on the uncorrelated condition is given by
\begin{eqnarray}
	\bar{F^{e}}=\displaystyle\frac{1}{M}\sum_{i=1}^{M}
\hbox{Tr}\biggl[ P^{+}_{_{RS_{i}}}\   
\frac{X _{_{RS_{k}}}\otimes\bigotimes _{l\ne k}^{M} Y _{_{S_{l}}}^{l}}{\hbox{Tr}\lbrack X _{_{RS_{k}}} \otimes\bigotimes _{l\ne k}^{M} Y _{_{S_{l}}}^{l}\rbrack }\biggr] \nonumber\\
=\displaystyle\frac{1}{M}\sum_{i=1}^{M}
\hbox{Tr}\biggl[ P^{+}_{_{RS_{i}}}\ \biggl( \tilde{X} _{_{RS_{k}}}\otimes
\bigotimes _{l\ne k}^{M} \tilde{Y} _{_{S_{l}}}^{l}\biggr)\biggr] 
\label{oefu}
\end{eqnarray}
for a certain $k$. Here
$\tilde{X} _{_{RS_{k}}}\equiv X _{_{RS_{k}}}/\hbox{Tr}\lbrack X _{_{RS_{k}}}\rbrack $
and $\tilde{Y} _{_{S_{l}}}^{l}\equiv Y _{_{S_{l}}}^{l}/\hbox{Tr}\lbrack Y _{_{S_{l}}}^{l}\rbrack$
are unit-trace (positive) operators.

\subsection{Probabilistic case}\label{pro}
Here we shall optimize $\bar{F^{e}}$ given in (\ref{oefu}) for a probabilistic channel.
Note that since the channel is not trace-preserving, 
there is no need to hold $A_{_{R}}=\1/d$. Firstly, we consider
the asymmetric case, where any conditions are not imposed on 
$A _{_{RS_{1}S_{2}\cdots S_{M}}}$ except for the condition (\ref{okk}).
Then Eq. (\ref{oefu}) implies that
all the entanglement fidelity $F^{e}_{s_{i}}$ except for $F^{e}_{s_{k}}$ can be written as
\begin{eqnarray}
F^{e}_{s_{i}}
=\displaystyle\frac{1}{d}\hbox{Tr}\biggl[ (\tilde{Y} ^{i})^{T}_{_{R}}\ \tilde{X} _{_{R}}\biggr]\ \ (i\ne k) .\label{oefi}
\end{eqnarray}
Here, we used the identity
$\lbrack O ^{T}_{_{R}}\otimes\1_{_{S_{i}}}\rbrack P^{+}_{_{RS_{i}}}=\lbrack\1_{_{R}}\otimes O _{_{S_{i}}}\rbrack\,P^{+}_{_{RS_{i}}}$ for an arbitrary operator $O$, where $T$ denotes the transposition.
Each $F^{e}_{s_{i}}$ for $i \ne k$ is obviously optimized when 
$(\tilde{Y} ^{i})^{T}=\vert\alpha\rangle\langle\alpha\vert$, 
where $\vert\alpha\rangle$ is the eigenstate with the maximum eigenvalue of $\tilde{X} _{_{R}}$.
These conditions for $i \ne k$ are all simultaneously satisfied, and hence we have 
\begin{eqnarray}
	\bar{F^{e}}&=&\displaystyle\frac{1}{M}\sum_{i=1}^{M}
\hbox{Tr}\biggl[ P^{+}_{_{RS_{i}}}\  
\biggl( \displaystyle\tilde{X} _{_{RS_{k}}}\otimes\bigotimes _{l\ne k}^{M} 
\vert\alpha^{\ast}_{_{S_{l}}}\rangle\langle\alpha^{\ast}_{_{S_{l}}}\vert \biggr)\biggr]  \label{oenk}\\
&=&\displaystyle\frac{1}{M}
\hbox{Tr}\biggl[ \Bigl\{ P^{+}_{_{RS_{k}}}+ \displaystyle\frac{M-1}{d} 
(\vert 0_{_{R}}\rangle\langle 0_{_{R}}\vert\otimes \1_{_{S_{k}}})
\Bigr\}\tilde{X}' _{_{RS_{k}}}\biggr],\nonumber
\end{eqnarray}
where $\tilde{X}' _{_{RS_{k}}}=(U_{_{R}}\otimes U^{\ast}_{_{S_{k}}})\tilde{X} _{_{RS_{k}}}(U^{\dagger}_{_{R}}\otimes U^{\ast\dagger}_{_{S_{k}}})$ and $U\vert\alpha\rangle =\vert 0\rangle$. Here, $U$ is a unitary operator.  
Therefore, we can optimize $\bar{F^{e}}$ by choosing 
$\tilde{X}' _{_{RS_{k}}}$
as the eigenstate with the maximum eigenvalue of $P^{+}_{_{RS_{k}}}+ (M-1)(\vert 0_{_{R}}\rangle\langle 0_{_{R}}\vert\otimes \1_{_{S_{k}}})/d 
$,
that is,
\begin{eqnarray}
\lambda _{m}=\displaystyle\frac{M+d-1+D(d, M)}{2d},\label{oemi}
\end{eqnarray}
where $D(d, M)=\sqrt{(M+d-1)^{2}-4(M-1)(d-1)}$. From the relation (\ref{paven}), we obtain 
the optimal average fidelity
\begin{eqnarray}
\bar{F}=\displaystyle\frac{3M+d-1+D(d, M)}{2M(d+1)},\label{ofm}
\end{eqnarray} 
and then the average fidelity for each output is given by
\begin{eqnarray}
F_{_{S_{k}}}&=&\displaystyle\frac{(2-d)\xi +d +2\sqrt{\xi (1-\xi )(d-1)}}{d+1}\label{ofmk} \\
F_{_{S_{i}}}&=&\displaystyle\frac{\xi +1}{d+1}\ \ \ \ \ (i\ne k),\label{ofmo}
\end{eqnarray} 
where $\xi =1/\lbrack 1+(d-1)/(d\lambda_{m}-d+1)^{2}\rbrack$.
The channel to realize these optimal values is given by, for instance, 
\begin{eqnarray}
	\vert\psi\rangle \underset{p_{\psi}}{\longrightarrow}
    \displaystyle\frac{M\vert\psi_{_{S_{k}}}\rangle}{\| M\vert\psi_{_{S_{k}}}\rangle\|}\otimes\bigotimes _{l\ne k}^{M} \vert 0_{_{S_{l}}}\rangle ,
	\label{pac}
\end{eqnarray}
where $M=\vert 0\rangle\langle 0\vert
+\gamma \sum_{i=1}^{d-1}\vert i\rangle\langle i\vert\ (\gamma =\sqrt{(1-\xi)/\xi (d-1)})$ and
$p_{\psi}$ is given by $p_{\psi}=\hbox{Tr} \bigl[ M^{\dagger} M \vert\psi\rangle\langle\psi\vert\bigr]$.
We can see that the corresponding optimal channel is realized by the two-valued measurement
whose positive-operator valued measure (POVM) elements are $\bigl\{\Pi_{0}= M^{\dagger}M,\ \Pi_{1}=\1-M^{\dagger}M\bigr\}$. Only when the outcome is $0$ (with the probability $p_{\psi}$),
the channel outputs $M\vert\psi_{_{S_{k}}}\rangle/\| M\vert\psi_{_{S_{k}}}\rangle\|$
on $S_{k}$ and $\vert 0\rangle$ on the other systems.\\

We next impose the symmetric condition, which means that all the fidelities for outputs
take the same value. Noting that the second term of Eq. (\ref{oenk}) corresponds to the fidelities
except for that in $S_{k}$, the maximum of their fidelities obviously is equal to the maximum eigenvalue of
$(\vert 0_{_{R}}\rangle\langle 0_{_{R}}\vert\otimes \1_{_{S_{k}}})/d$.
That is $F^{e}_{s_{i}}\le 1/d\ (i\ne k)$.
Moreover, in fact, if choosing
$\tilde{X}' _{_{RS_{k}}}$ as the eigenvector with the maximum 
eigenvalue of $(\vert 0_{_{R}}\rangle\langle 0_{_{R}}\vert\otimes \1_{_{S_{k}}})/d$,
we realize that it also gives $F^{e}_{s_{k}}= 1/d$ from Eq. (\ref{oenk}).
Then, that choice satisfies the symmetric condition, so
we obtain the optimal average fidelity 
\begin{eqnarray}
\bar{F}=\displaystyle\frac{2}{d+1}.\label{ofms}
\end{eqnarray} 
The corresponding channel is then given by, for instance,
\begin{eqnarray}
	\vert\psi\rangle 
	\ \ \underset{p_{\psi}}{\longrightarrow}\ \ 
	\bigotimes _{i=1}^{M} \vert \alpha _{_{S_{i}}}\rangle.\label{psc}
\end{eqnarray}
Here, $p_{\psi}$ is given by $p_{\psi}=\vert\langle\alpha\vert\psi\rangle\vert^{2}$.
This channel is also realized by the two-valued measurement whose POVM elements are 
$\bigl\{ \Pi_{0}=\vert \alpha\rangle\langle \alpha\vert,\ \Pi_{1}=\1-\vert \alpha\rangle\langle \alpha\vert\bigr\}$. This channel always outputs the result $\vert\alpha\rangle$ for all the systems $S_{i}$ with the probability $p_{\psi}$.

\subsection{Deterministic case}\label{det}
Here, we consider the deterministic channels. Since the deterministic condition can be written as  
$A_{_{R}}=\1/d$, by imposing this condition on Eq. (\ref{oefi}), it follows that
all the entanglement fidelity $F^{e}_{s_{i}}$
except for $F^{e}_{s_{k}}$ must be fixed to
\begin{eqnarray}
F^{e}_{s_{i}}=\displaystyle\frac{1}{d^{2}}\ \ \ \ (i\ne k). \label{oeda}
\end{eqnarray}
Then, the entanglement fidelity $F^{e}_{s_{k}}$ can be obviously maximized by 
choosing $X _{_{RS_{k}}}=P^{+}_{_{RS_{k}}}$ ({\it i.e., }$F^{e}_{s_{k}}=1$) 
satisfying the deterministic condition.
Therefore the optimal average fidelity is
\begin{eqnarray}
\bar{F}=\displaystyle\frac{1}{M}+\frac{M-1}{dM},\label{ofd}
\end{eqnarray}  
and the corresponding channel is given by, for instance,
\begin{equation}
	\vert\psi\rangle\rightarrow
	\vert\psi_{_{S_{k}}}\rangle\otimes\bigotimes _{l\ne k}^{M}\frac{\1_{_{S_{l}}}}{d}.\label{dac}
\end{equation}

In the deterministic and symmetric channel, the entanglement fidelity
$F^{e}_{s_{k}}$ also has to be $1/d^{2}$ 
because the other entanglement fidelities $F^{e}_{s_{i}}$ are fixed to $1/d^{2}$
by the deterministic condition. Therefore the optimal average fidelity is 
\begin{eqnarray}
\bar{F}=\displaystyle\frac{1}{d},\label{ofds}
\end{eqnarray}
and the corresponding channel is given by, for instance,
\begin{equation}
	\vert\psi\rangle\rightarrow
	\bigotimes _{i=1}^{M}\frac{\1_{_{S_{i}}}}{d}.\label{dsc}
\end{equation}
\begin{table*}
	\begin{center}
	{\renewcommand\arraystretch{2.0}
    \small
	\begin{tabular}{|c||c|c|}\hline
	         & Symmetric   & Asymmetric            \\ \hline\hline
	Deterministic & $\displaystyle\frac{1}{d}$\ $(F_{_{S_{k}}}=F_{_{S_{i}}}=\frac{1}{d})$
    &$\displaystyle\frac{1}{M}+\frac{M-1}{dM}$\ \ \ ($F_{_{S_{k}}}=1, F_{_{S_{i}}}=\frac{1}{d}$)
          \\ \hline
	Probabilistic & $\displaystyle\frac{2}{d+1}$\ ($F_{_{S_{k}}}=F_{_{S_{i}}}=\frac{2}{d+1}$) 
@@& $\displaystyle\frac{3M+d-1+D_{_{M}}}{2M(d+1)}$
    ($F_{_{S_{k}}}=\frac{(2-d)\xi +d +2\sqrt{\xi (1-\xi )(d-1)}}{d+1}, F_{_{S_{i}}}=\frac{\xi +1}{d+1} $) \\ \hline
	\end{tabular}}
	\caption{\ \ \ \ \ \ The optimal average fidelities in the $d$-dimensional $1\rightarrow M$ uncorrelated channels. The average fidelity for each output is also shown in the bracket.}\label{tau}
	\end{center}
\end{table*}

\section{Conclusion}\label{con}
The derived optimal fidelities for all four cases are summarized in TABLE \ref{tau}.
We notice
that the optimal fidelities in the uncorrelated channels remarkably vary according to
the conditions imposed on the channels. 
This result seems to present the striking contrast to the imperfect cloning
where the optimal fidelity does not change at all and is $\bar{F}=5/6$ for every
condition in $d=2$, for instance (since the value $\bar{F}=5/6$ is known to coincide with
the boundary of the no-signaling condition \cite{gi}, such an invariance is expected).
 
As we have seen in the previous section,
the optimal deterministic and asymmetric uncorrelated channel can be realized by attaching randomized
states to the intact input state, where the optimal average fidelity is thus
$1/M+(M-1)/dM$ (this optimal channel is the same as the one called {\it trivial amplification}
in Ref. \cite{scib}),
and the optimal deterministic and symmetric channel can be realized by
randomizing all output states, where the optimal average fidelity is $1/d$. 
In these cases (i.e., in the deterministic ones), 
the fidelities at multiple outputs cannot exceed
$1/d$, that is, {\it quantum-information division} is impossible.
This impossibility is also expected from \cite{dade}.

On the other hand, 
in the optimal probabilistic uncorrelated channels, all
the average fidelities at output can exceed $1/d$
(even in the symmetric channel). It is interesting that, even in this case,
the optimal asymmetric channel is realized 
by attaching randomized states to an input-dependent state, and the optimal symmetric one is 
realized by randomizing all output states. With this similarity to the deterministic case, however, 
in the probabilistic channels
whether the outputs exist or not can contain the input information, through
which the randomized states can indirectly depend on the input states.
This is why all the average fidelities at output can exceed $1/d$.
Therefore, we can conclude that {\it quantum-information division} is possible probabilistically.

\section*{Acknowledgements}
We thank N. Hatakenaka for helpful comments.
This work was supported by JSPS KAKENHI Grants No.
23246071 and No. 24540405.

\appendix
\section*{Appendix}
Here we shall derive the state $A$ isomorphic to an uncorrelated channel $\Lambda$. 
Noting that the output state of $\Lambda$ for the pure state input $\vert \psi\rangle$ is given by
$ d\langle\psi _{_{R}}^{\ast}\vert A _{_{RS_{1}S_{2}\cdots S_{M}}}
	\vert\psi _{_{R}}^{\ast}\rangle$ and that it is impossible to clone the input state 
$\vert \psi\rangle$ to multiple outputs without correlation (even if imperfectly) \cite{dade},
the condition of an uncorrelated channel becomes
\begin{eqnarray}
	\langle\psi _{_{R}}^{\ast}\vert A _{_{RS_{1}S_{2}\cdots S_{M}}}
	\vert\psi _{_{R}}^{\ast}\rangle\ \propto
\rho_{s_{_{k}}}(\psi )\otimes\bigotimes _{l\ne k}^{M}\rho_{s_{_{l}}}\label{apun}
\end{eqnarray} 
for any $\vert\psi_{_{R}}\rangle$, where $\rho_{s_{k}}(\psi )$ is a state
depending on $\vert\psi\rangle$ in the $k$-th output and $\rho_{s_{_{l}}}$ are fixed 
output states (not depending on $\vert \psi\rangle$). 

For our derivation, we need to prove the following lemmas:

{\it Lemma 1.} For any operator $W_{AB}$ in the system $A$ and $B$, 
it has a product form $W_{AB}=W_{A}\otimes W_{B}$ if and only if
\begin{eqnarray}
    \langle \zeta _{_{A}}\vert W_{AB}\vert \eta _{_{A}}\rangle \propto W_{B}\label{apra}
\end{eqnarray} 
for all $\vert \zeta_{_{A}}\rangle$ and $\vert \eta_{_{A}}\rangle$. Here $\{\vert \zeta\rangle\}$
and $\{\vert \eta\rangle\}$ are complete orthonormal basis in the system $A$.

{\it Proof}. It is trivial that the condition (\ref{apra}) is necessary, so we shall prove the
sufficiency.
Note that the operator $W_{AB}$ can be decomposed with complete orthonormal basis as 
	$W_{AB}=\sum_{\zeta\eta ij}
    c_{_{\zeta\eta ij}}\vert \zeta_{_{A}}\rangle\langle \eta_{_{A}}\vert\otimes	
    \vert i_{_{B}}\rangle\langle j_{_{B}}\vert$.
Therefore, 
from the assumption, we can write $\sum_{ij}
    c_{_{\zeta\eta ij}}\vert i_{_{B}}\rangle\langle j_{_{B}}\vert=a_{_{\zeta\eta}} W_{B}$
by means of some coefficient 
$a_{\zeta\eta}$. Thus we obtain
\begin{eqnarray}
	W_{AB}=\displaystyle\sum_{\zeta\eta }a_{_{\zeta\eta}}
    \vert \zeta_{_{A}}\rangle\langle \eta_{_{A}}\vert\otimes	
    W_{B} . \ \square \label{aprb}
\end{eqnarray}

{\it Lemma 2.} The output state has the form 
\begin{eqnarray}
	\langle\psi _{_{R}}^{\ast}\vert A _{_{RS_{1}S_{2}\cdots S_{M}}}
	\vert\psi _{_{R}}^{\ast}\rangle\ \propto
\rho_{s_{_{k}}}(\psi )\otimes\rho_{\bar{s}_{_{k}}}\label{aprc}
\end{eqnarray} 
for any $\vert\psi\rangle_{_{R}}$, where $\bar{S}_{k}$ denotes all output systems except for the $k$-th output system, if and only if 
\begin{eqnarray}
	\langle k _{_{R}}\vert\langle m _{_{S_{_{k}}}}\vert A _{_{RS_{1}\cdots S_{M}}}\vert l _{_{R}}\rangle\vert n _{_{S_{_{k}}}}\rangle \propto \hbox{Tr}_{_{RS_{k}}}\lbrack A _{_{RS_{1}\cdots S_{M}}}\rbrack\label{aprd}
\end{eqnarray} 
holds for all $\vert k_{_{R}}\rangle$, $\vert l_{_{R}}\rangle$, 
$\vert m _{_{S_{_{k}}}}\rangle$, and $\vert n _{_{S_{_{k}}}}\rangle$. Here $\{\vert k\rangle\}$
is a complete orthonormal basis in the system $R$ or $S_{_{k}}$.

{\it Proof.} By the condition (\ref{aprc}), the relation
\begin{eqnarray}
	\langle \psi _{_{R}}^{\ast}\vert\langle m _{_{S_{_{k}}}}\vert A _{_{RS_{1}S_{2}\cdots S_{M}}}\vert \psi _{_{R}}^{\ast}\rangle\vert n _{_{S_{_{k}}}}\rangle \propto \rho_{\bar{s}_{_{k}}}\nonumber
\end{eqnarray} 
holds for all $\vert m _{_{S_{_{k}}}}\rangle$ and $\vert n _{_{S_{_{k}}}}\rangle$, and any $\vert\psi_{_{R}}\rangle$.
Taking $\vert\psi _{_{R}}^{\ast}\rangle$ as $\vert k_{_{R}}\rangle$, $\vert l_{_{R}}\rangle$,
$(\vert k_{_{R}}\rangle +\vert l_{_{R}}\rangle)/\sqrt{2}$, and
$(\vert k_{_{R}}\rangle +i\vert l_{_{R}}\rangle)/\sqrt{2}$, it follows that the relation
\begin{eqnarray}
\langle k _{_{R}}\vert\langle m _{_{S_{_{k}}}}\vert A _{_{RS_{1}S_{2}\cdots S_{M}}}\vert l _{_{R}}\rangle\vert n _{_{S_{_{k}}}}\rangle \propto \rho_{\bar{s}_{_{k}}}\nonumber
\end{eqnarray} 
holds for all $\vert k_{_{R}}\rangle$, $\vert l_{_{R}}\rangle$, 
$\vert m _{_{S_{_{k}}}}\rangle$, and $\vert n _{_{S_{_{k}}}}\rangle$. And by noting $\rho_{\bar{s}_{_{k}}}$
is a fixed state and the definition of trace, $\rho_{\bar{s}_{_{k}}}\propto\hbox{Tr}_{_{RS_{k}}}\lbrack A _{_{RS_{1}S_{2}\cdots S_{M}}}\rbrack$ holds. Conversely, we assume Eq. (\ref{aprd}). If choosing $\vert k_{_{R}}\rangle=\vert l_{_{R}}\rangle=\vert\psi _{_{R}}^{\ast}\rangle$, it becomes
\begin{eqnarray}
	\langle \psi _{_{R}}^{\ast}\vert\langle m _{_{S_{_{k}}}}\vert A _{_{RS_{1}S_{2}\cdots S_{M}}}\vert \psi _{_{R}}^{\ast}\rangle\vert n _{_{S_{_{k}}}}\rangle \propto \hbox{Tr}_{_{RS_{k}}}\lbrack A _{_{RS_{1}S_{2}\cdots S_{M}}}\rbrack\nonumber
\end{eqnarray} 
for all $\vert m _{_{S_{_{k}}}}\rangle$, $\vert n _{_{S_{_{k}}}}\rangle$, and $\vert\psi_{_{R}}\rangle$. Here
applying Lemma 1 to this equation, we obtain
\begin{eqnarray}
	\langle\psi _{_{R}}^{\ast}\vert A _{_{RS_{1}S_{2}\cdots S_{M}}}
	\vert\psi _{_{R}}^{\ast}\rangle\ \propto
\rho_{s_{_{k}}}(\psi )\otimes\hbox{Tr}_{_{RS_{k}}}\lbrack A _{_{RS_{1}S_{2}\cdots S_{M}}}\rbrack. \square
\nonumber
\end{eqnarray} 

{\it Corollary.} The output state has the form of (\ref{aprc}) if and only if the isomorphic state has the form of
\begin{eqnarray}
 A _{_{RS_{1}S_{2}\cdots S_{M}}}=X _{_{RS_{k}}} \otimes Y _{_{\bar{S}_{k}}}. \label{apri}
\end{eqnarray}

{\it Proof.} By Lemma 2, we may assume Eq. (\ref{aprd}).
 Since Eq.(\ref{aprd}) holds for all $\vert k_{_{R}}\rangle$, $\vert l_{_{R}}\rangle$, 
$\vert m _{_{S_{_{k}}}}\rangle$, and $\vert n _{_{S_{_{k}}}}\rangle$, it is equivalent to hold
\begin{eqnarray}
	\langle \Psi _{_{RS_{_{k}}}}\vert A _{_{RS_{1}S_{2}\cdots S_{M}}}\vert \Psi' _{_{RS_{_{k}}}}\rangle \propto \hbox{Tr}_{_{RS_{k}}}\lbrack A _{_{RS_{1}S_{2}\cdots S_{M}}}\rbrack\label{aprj}
\end{eqnarray} 
for any $\vert \Psi _{_{RS_{_{k}}}}\rangle $ and $\vert \Psi' _{_{RS_{_{k}}}}\rangle $. So applying Lemma 1, it follows that the above is
equivalent to the isomorphic state having the form of (\ref{apri}).
The converse is trivial.$\square$

Finally, since in our case we are focusing on the output state in the system $\bar{S}_{k}$ as
$\rho_{\bar{s}_{_{k}}}=\bigotimes _{l\ne k}^{M}\rho_{s_{_{l}}}$, the corresponding isomorphic state  must obviously have the form of 
\begin{eqnarray}
 A _{_{RS_{1}S_{2}\cdots S_{M}}}=X _{_{RS_{k}}} \otimes\bigotimes _{l\ne k}^{M} Y _{_{S_{l}}}^{l}. \label{aprk}
\end{eqnarray}

\end{document}